\documentclass[mathleft,fleqn,%
]{an}
\usepackage{graphicx}
\usepackage[varg]{txfonts}
\begin{document}

\Pagespan{1}{}
\Yearpublication{2014}%
\Yearsubmission{2014}%
\Month{0}%
\Volume{999}%
\Issue{0}%
\DOI{asna.201400000}%

\title{New Aspects of Photon Propagation in Expanding Universes}

\author{H.-J. Fahr\inst{1}\fnmsep\thanks{Corresponding author:
        {hfahr@astro.uni-bonn.de}}
\and  M. Heyl\inst{2}
}
\titlerunning{New Aspects of Photon Propagation in Expanding Universes}
\authorrunning{H.-J. Fahr \& M. Heyl}
\institute{
Argelander Institut f{\"u}r Astronomie, Universit{\"a}t Bonn\\ Auf dem H{\"u}gel 71, 53121 Bonn, Germany
\and 
Deutsches Zentrum f{\"u}r Luft und Raumfahrt (DLR)\\ K{\"o}nigswinterer Str. 522 - 524, 53227 Bonn, Germany}

\received{XXXX}
\accepted{by Astronomische Nachrichten in August 2016}
\publonline{XXXX}


\abstract{
  According to present cosmological views the energy density of CMB (Cosmic
Microwave Background) photons, freely propagating through the expanding
cosmos, varies proportional to $1/S^{4}$ with $S$ being the scale factor of
the universe. This behavior is expected, because General Theory of
Relativity, in application to FLRW- (Friedmann-Lemaitre-Robertson-Walker)
cosmological universes, leads to the conclusion that the photon wavelengths
increase during their free passage through the spacetime metrics of the
universe by the same factor as does the scale factor $S$. This appears to be
a reasonable explanation for the presently observed Planckian CMB spectrum with its actual
temperature of about 2.7\,K, while at the time of its origin after the
last scattering during the recombination phase its temperature should have
been about 3000\,K, at an epoch of about 380\,ky after the Big Bang,
when the scale of the universe $S_{\mathrm{r}}$ was smaller by roughly a factor of $
S/S_{\mathrm{r}}=1+z_{\mathrm{r}}=1100$ compared to the present scale $S=S_{0}$ of the universe.
In this paper we start from putting the question whether the scale-behavior of the 
CMB energy density that enters the energy-momentum tensor of the field equations 
describing the expanding universe is really falling off like $S^{-4}$ and, if 
in fact a deviation from a behavior according to $S^{-4}$ would occur, why do we 
nevertheless presently observe a CMB energy density which appears to be in accordance 
with such a $S^{-4}$-scaling? This question arises from another basic and 
fundamental question, namely: Can we really assume that the wavelength of the 
freely propagating photon during its travel all the way along its light geodetic is 
permanently affected by the expansion of the universe, i.e continuously  recognizes the 
expansion of the cosmic scale $S$? With other words: Do freely propagating photons 
really undergo a permanent change of their wavelengths when freely traveling through 
cosmic space-time or is the observationally apparent energy loss of cosmologically 
red-shifted photons an effect which only occurs just in the moment of photon 
registration at some specific world point?
If the latter would prove to be true, then it would mean that the energy density of 
freely propagating, non-interacting CMB photons, due to non-changing, conserved wavelengths, is 
behaving with respect to cosmic scale variation different from conventional expectations, 
but rather would turn out to behave just like the energy density of matter, namely 
according to $S^{-3}$. Hence the photon part of the energy momentum tensor would become
different and associated solutions of FLRW-equations would undergo corresponding 
modifications. In consequence, the CMB energy density as far as it enters the energy-momentum tensor 
generated by freely propagating CMB photons during the expansion period of the universe after the recombination 
era would no longer become negligible for the cosmic dynamics, since its value 
would stay in the same order of magnitude as that of baryonic or dark matter.}

\maketitle

\section{Introduction}
In the following considerations we start from the basis both
of Einstein's Special and General Theory of Relativity (STR and GTR) where
it is stated that for a photon moving with the velocity of light $c$ the
so-called proper time increment (time increment $d\tau _{0}$ in the photon
proper system) vanishes, i.e. $d\tau _{0}=0$ (see e.g. Rindler 1977; 
Tolman 1987; Misner et al. 1973; Goenner 1996). If this statement needs
to be taken serious, then one seriously has to discuss the consequences of
it and the conclusions to be drawn there from. Obviously it would mean that
a photon moving along a geodetic trajectory (i.e. along its light geodetic)
from spacepoint A to spacepoint B in its own reference system does not
require proper time to cover this passage. Of course this should be true
only in its individual reference system. In other words: the photon starts
at some worldtime $t_\mathrm{{A}}$ in\ a spacepoint A and arrives later at a
spacepoint B at a worldtime $t_\mathrm{{B}}$, however, this all happens while in
the individual proper frame of this photon no time has lapsed, i.e. the
light geodetic is an isochronal line for the individual photon. 

The question then arises: How can a photon, moving freely in a cosmological spacetime,
react to the change of the dynamical spacetime geometry, if in its proper
system no time lapses? Should not its physical state be strictly preserved
in case of no lapse of time? We presume that due to the lack of proper time
lapse the photon will simply not have the chance to react to dynamical
spacetime geometry in the generally expected form, because without lapse of
time nothing, i.e. no change can happen in the photons` reference frame. The consequences of
that presumption, if confirmed as true, are enormous, meaning for instance
that a photon freely propagating in cosmological spacetime, i.e. in a free falling system does not change
its proper state, i.e. its energy and spin and, when seen in the wave
picture of the photon, also its wavelength. In the photon frame no change of
the photon state should happen, neither due to a change of an environmental
gravitational potential, nor due to the expansion of the universe, i.e. any
change of the cosmological metrics. The photon at free propagation simply
would preserve its physical state which it got at the moment of its
generation from an electromagnetic event, like emission from a stellar
surface or from the CMB horizon. The so-called and generally believed
cosmological redshift of its wavelength in an expanding universe 
(Misner et al. 1973; Berry 1990; Sexl \& Urbantke 1987; Peacock 1999; Pecker \& Vigier 1987) would thus simply have to
be caused by the cosmic observer that measures the photon frequency with its
local world clock, comparable perhaps with the special relativistic Doppler
shift which results from the observer's relative motion with respect to the
photon frame, i.e. the effect of the observer.

In the following sections of this paper we would like to
explain more in detail the physical arguments behind the above idea. We then
calculate the effects of the proposed constant wavelength of freely
propagating photons (we call them \textbf{"free photons"} in the rest of the paper) 
in the cosmic era after recombination $(z\leq z_{\mathrm{r}} \approx 1100)$.
Finally we discuss the implications that are connected with non-redshifted
photons entering the energy momentum tensor $T_{\mathrm{ik}}$ of Robertson-Walker universes
and resulting cosmological expansion solutions.

\section{Theory of the cosmological redshift}
\subsection{Change of the beat of global worldtime}
In a universe with Robertson-Walker metric the world line element $dl$ is
given by (see Goenner 1996)

\begin{equation}
dl^{2}=c^{2}dt^{2}-S^{2}(t)[\frac{dr}{1-kr^{2}}-r^{2}d\Omega ^{2}]. \label{1}
\end{equation}

As is well known from SRT, but valid also in GRT, for the wordline of a
photon, a so-called light-geodetic, one has

\begin{equation}
dl^{2}=c^{2}d\tau_{0}^{2}=0 \label{2}
\end{equation}

meaning that in the proper system of the photon the proper time increment $%
d\tau _{0}$ vanishes, implying that while the photon is propagating through
space in its reference system no proper time is running, i.e. the photon in
its proper system does not age, i.e. it consequently has to conserve its properties. That
obviously must mean that during the passage of such a photon through cosmic
spacetime this photon cannot and does not change its identity, though this
photon on its light geodetic reaches worldpoints of different world ages.
This implies that the photon does not change its physical signatures or
characteristics like e.g. its energy or its angular momentum, i.e. during its
free flight along light geodetics it conserves those properties with which
it was generated. This is because an object, as also a photon in the quantum
duality view seen as a particle, cannot change its physical signatures under
any physical action\ which does not have a finite increment in time $d\tau
_{0}$ to act.

\subsection{Beat of the time in a static gravitational field}
To give more credit to this view we would like to mention the following aspect: 
The above argumentation is well familiar to SRT scientists, is standardly used
to carry out time synchronization at different places and can easily be
compared and controlled with the standard practice of synchronizing times in
a Schwarzschildian spacetime geometry. 

We shall follow here the presentation given by \textit{Rindler 1977} and shall
first pick up the most important idea presented there. Assuming that the
gravitational field of the central Schwarzschildian mass acts on a photon
freely propagating in the predetermined spacetime geometry then allows to
describe the photon propagation as happening in a free falling, inertial system (photon
in a free falling cabin). As \textit{Rindler 1977} clearly and correctly states, in
that free falling cabin (inertial system = ff-system) no change in frequency of the
propagating photon occurs, i.e. the free falling photon does not change its
frequency in this system. The question only is how this fact can be
transfered into the non-inertial system, i.e. into the fixed Scharzschildian
coordinate system. The argumentation then is that the free-falling system
(cabin) has first to be replaced by an equivalent inertial system
calculating the velocity $du$ that the free fall over a propagation distance 
$dl$ induces due to the acting gravitational acceleration $g=d\Phi /dr$.
Neglecting second order corrections one finds that the equivalent velocity
increase \ is given by

\begin{equation}
du=(dl/c)g=(dl/c)(d\Phi /dr). \label{3} 
\end{equation}

Transforming now from the equivalent inertial system (ff-system) into the fixed
Schwarzschildian system (ss-system) needs the SRT\ frequency transformation
from the system moving with $du$ into the system at rest,i.e. leading to

\begin{eqnarray}
\nu _{\mathrm{ss}}=\nu \sqrt{(1+du/c)/(1-du/c)} \simeq \nonumber 
\\ \nu \sqrt{(1+du/c)^{2}}=\nu \label{4}
(1+du/c)
\end{eqnarray} 

or 

\begin{equation}
\nu _{\mathrm{ss}}-\nu =\nu (du/c) \label{5}
\end{equation}

resulting in 

\begin{equation}
d\nu /\nu=(\nu _{\mathrm{ss}}-\nu )/\nu =(du/c) \label{6}
\end{equation}

leading with  $d\nu /\nu=d\ln \nu $ and $dl\cdot g/c=-dr\cdot grad\Phi
/c=d\Phi $ to the well known result that 

\begin{eqnarray}
\nu _{ss}=\nu \cdot \exp [\int (dl/c^{2})(d\Phi /dr)= \nonumber
\\ \nu \cdot \exp [-\frac{1}{c^{2}}\int_{\Phi _{1}}^{\Phi _{2}}d\Phi]= \nu \cdot \exp [-\frac{\Delta\Phi _{1,2}}{c^{2}}] \label{7}
\end{eqnarray}

That finally again expresses, also in the case of this Scharzschildian spacetime
metric, the remarkable fact that a photon, - not while freely propagating
over a region connected by a potential difference $\Delta \Phi _{1,2}$ -,
but only, when measured in the static Schwarzschild frame, reflects a
frequency shift of the above mentioned order. 

This context is commonly interpreted as saying: having two clocks at
different positions $r_{1}$ and $r_{2}$ within a Schwarzschild metric, one
must expect a difference in the beats of the time $t_{1}$ and $t_{2}$ at
these different positions. Since photons on spacepoint connecting geodetics
are perfect time-synchronizers in spacetime, they conserve the time digits.
If thus a photon transports time signatures connected with its frequency $%
\nu _{1}$at $r_{1}$ to the spacepoint $r_{2}$, then these time signatures
arriving at $r_{2}$ are registered there by the local beat of the time, i.e.
by

\begin{equation}
\nu _{2}=\nu _{1}\exp [-\frac{\Delta \Phi _{1,2}}{c^{2}}] \label{8}
\end{equation}

This phenomenon does not mean that the photon changes its frequency while
passing along its light geodetic, rather the photon is emitted with its
genuine frequency $\nu _{1}$ and keeps it in the ff--system, but when
registered at a place $r_{2}$ its frequency appears, as if it had locally
changed to the frequency $\nu _{2}$. This means the so-called Moesbauer
effect does not prove the change of the photon frequency in free geodetic
flights, but the new evaluation of the original frequency $\nu _{1}$ with a
new local clock detuned with respect to the one by which the photon
originally was emitted.

This is both very interesting and very hard to understand,
since on one hand with $d\tau _{0}=0$ the identity of the photon must be
preserved, while nevertheless this photon permanently reaches spacepoints
where the worldtime has changed $t\geq t_{0}$, i.e. the non-aging photon
along its trajectory meets spacepoints of a globally aging universe.

\subsection{Photons in their own rest frame}
Hence we want to ask now, why the well-founded and
well-believed phenomenon of a cosmological redshifting of a freely
propagating photon can occur at all, if on the other hand the photon cannot
change its physical state due to the lack of any time lapse in its ff-system. For that we go
back to the line element of the freely propagating photon in a given dynamic
cosmic spacetime geometry within the frame of the Friedmann-Lema\^{i}tre-Robertson-Walker cosmology. 
With $d\tau _{0}=d\Omega =0$ we find from Eq. (\ref{1}):

\begin{equation}
\xi (r)=\int_{r_{0}}^{r}\frac{dr}{\sqrt{1-kr^{2}}}=-c\int_{t_{0}}^{t}\frac{dt}{S(t)}. \label{9}
\end{equation}

Now it is generally assumed that a photon is an extended
wave, allowing to write the analogue of the above equation not only for the
beginning of the wavetrain (phase "a") , but also for the end of the
wavetrain (phase "b"). Adopting then that during the time $\delta
t_{0}=\lambda _{0}/c$ at the place of the photon emission the geometry of
the universe has not changed, will then permit to write the following
equation for the propagation of phase "b" of this photon

\begin{equation}
\xi (r)=\int_{r_{0}}^{r}\frac{dr}{\sqrt{1-kr^{2}}}=-c\int_{t_{0}+\delta \label{10}
t_{0}}^{t+\delta t}\frac{dt}{S(t)}
\end{equation}

and one obtains as the difference of these two equations

\begin{equation}
\frac{\delta t}{S(t)}=\frac{\delta t_{0}}{S(t_{0})} \label{11}
\end{equation}

or

\begin{equation}
\frac{\delta t}{\delta t_{0}}=\frac{\lambda }{\lambda _{0}}=\frac{S(t)}{S(t_{0})} \label{12}
\end{equation}

meaning that the wavelength of the photon during its
propagation has been changed according to the change in the universal scale $S(t)$ on its way (see e.g. Dodelson 2003; Harrison 2000). How to understand this?

If one says that the photon is characterized by an
wavelength $\lambda_{0}$ which it keeps during all its free
propagation, then it simply means the time increment $\delta t_{0}=\lambda_{0}/c
$ is different from that $\delta t=\lambda/c$, because the time
propagation rate (the beat of the time) has changed during the change in the
world time between time $t_{0}$ and time $t$.

That can be interpreted as if the photon on its
passage through cosmic spacetime keeps its identity, i.e. its wavelength $%
\lambda_{0}$,but when it passes over a detector at some spacepoint at
worldtime\ $t$, then it needs a time

\begin{equation}
\frac{\lambda }{c}=\frac{\lambda_{0}}{c}\frac{S(t)}{S(t_{0})}. \label{13}
\end{equation}

Assuming a constant velocity of light this can be interpreted
as saying that the beat of the global worldtime has changed between the
events of emission and absorption of the photon.

\subsection{Alternative view: Redshift as accumulated Doppler-Shift}
An alternative way to explain the expansion related 
redshift is to regard it as an accumulation of infinitesimal Doppler-Shifts along the 
photon's trajectory (Bunn \& Hogg 2008; Peacock 2008). The basic idea here is that the fractional frequency shift 
$\delta\nu/\nu$ for a photon which travels a distance $d=c\delta t$ within a time increment $\delta t$ 
is given by

\begin{equation}
\delta\nu/\nu = \delta v/c = -(Hd)/c = -H\delta t \label{14}
\end{equation}

with $H$ the Hubble parameter and $v$ the recession velocity of the photon emitting galaxy ($v<<c$). 
Now, using $H=\dot{S}/S$ with $S$ the scale parameter, we finally get

\begin{equation}
\delta\nu/\nu = -\delta S/S, \label{15}
\end{equation}

the integration of which again leads to the well-known redshift equation

\begin{equation}
\lambda = \lambda_0 S(t)/S(t_0). \label{16}
\end{equation}

It should be pointed out that a characteristic of the Doppler-Shift is a non-changing frequency/wavelength during 
the photon's travel from the emitter to the observer, no matter whether the relative velocity between emitter and observer is constant or not.  
This means that also the cosmological redshift - here seen as an accumulated Doppler-Shift - is just an effect 
which occurs in the moment of observation. During the travel of the photon its 
frequency/wavelength remains unchanged.

\subsection{The beat of the cosmic clock}
Starting from the first of the two Friedman-Lema\^{i}tre Robertson Walker
equations given in the form (see e.g. Goenner, 1996)

\begin{equation}
\left( \frac{\dot{R}(t)}{R(t)}\right) ^{2}=\frac{8\pi G}{3}\rho (t)-\frac{%
kc^{2}}{R^{2}(t)}+\frac{\Lambda c^{2}}{3}, \label{17}
\end{equation}

where $R$ and $\dot{R}$ are the scale of the universe and its time
derivative, respectively, $G$ is the gravitational constant, $\rho (t)$
denotes the cosmic matter density, $k$ is the curvature parameter, and $%
\Lambda $ denotes the cosmological parameter or the vacuum energy density,
respectively. As one can easily show (see e.g. Fahr \& Heyl, 2007) in a
homogeneous universe with vanishing curvature (i.e. $k=0!$) and vanishing
cosmologic parameter (i.e. $\Lambda =0!)$ one can formally, and in strict
analogy to classic mechanics, introduce a cosmic gravitational potential $%
\Phi _{\mathrm{cos}}$ which then is derived from this first of the two F-L-R-W
equations in the form:

\begin{equation}
\Phi _{\mathrm{cos}}(R)=\frac{8\pi G}{3}\rho R^{2}. \label{18}
\end{equation}

Assuming that no dark energy has to be taken into account (i.e. $\Lambda =0!$%
), and that only conservative forms of matter (i.e. massive particles with
conserved total numbers; no decay or creation of matter!) prevail, then the
matter density $\rho $ scales inversely proportional to the  cosmic volume,
hence with $R^{-3}$, and the above potential consequently can be re-written
in the form

\begin{equation}
\Phi _{\mathrm{cos}}(R)=\frac{8\pi G}{3}\frac{M}{R}, \label{19}
\end{equation}

where $M$ has been introduced as the total and constant mass of the
universe. The above expression then shows that in an expanding universe the
value of its local cosmic potential $\Phi _{\mathrm{cos}}(R)$ permantly decreases,
meaning that for $R,t\rightarrow \infty $ it asymptotically approaches the
value $\Phi _{\mathrm{cos}}(R\rightarrow \infty )=0$.

In comoving cosmological systems this situation can be compared with that of
clocks at different positions in a stationary central gravitational field
(see section 2.2). As is well known the beat of the clocks at different
positions of a gravitational potential in this case is detuned (e.g. see
Rindler, 1977). Analogously based, however, on the above cosmic potential,
the local cosmic beat of the time should be subject to the following
time-tuning law

\begin{equation}
t=t_{0}exp[-\frac{\Phi (R)}{c^{2}}]. \label{20}
\end{equation}

From the above one derives the following relation between frequencies of
CMB\ photons $\nu _{\mathrm{rec}}$ near the recombination period, when the scale of
the universe was $R=R_{\mathrm{rec}}$ , and the associated frequencies $\nu $ of such
CMB photons registered by detectors at the present time:

\begin{eqnarray}
\frac{\nu -\nu _{\mathrm{rec}}}{\nu }=\frac{\Delta \Phi _{\mathrm{eff}}}{c^{2}}=
-\frac{8\pi G}{c^{2}}[\rho _{\mathrm{rec}}R_{\mathrm{rec}}^{2}-\rho R^{2}]= \nonumber
\\ -\frac{8\pi GM}{c^{2}}[\frac{1}{R_{\mathrm{rec}}}-\frac{1}{R}], \label{21}
\end{eqnarray}

where in the upper relation the potential energy $m_{\nu }\Delta \Phi _{\mathrm{eff}}$
has been considered as small with respect to the photon energy itself $%
m_{\nu }c^{2}$. Neglecting the small terms in the brackets on LHS\ and RHS,
one then can find:

\begin{equation}
\frac{\nu _{\mathrm{rec}}}{\nu }=\frac{8\pi GM}{c^{2}R_{\mathrm{rec}}}. \label{22}
\end{equation}

According to the above relation, the unchanged frequency of the CMB\ photon
from the recombination phase would then, when registered today by a local
clock (i.e. spectrometer), appear with a frequency $\nu $ given by

\begin{equation}
\nu =\nu _{\mathrm{rec}}\frac{c^{2}R_{\mathrm{rec}}}{8\pi GM}. \label{23}
\end{equation}

Adopting at the time of cosmic matter recombination a Planck spectrum for
the CMB\ photons, one would have a spectral photon energy density at some
frequency $\nu _{1,\mathrm{rec}}$ in the Rayleigh-Jeans branch given by (e.g. see
Goenner, 1996):

\begin{eqnarray}
\varepsilon (\nu _{1,\mathrm{rec}})\sim \frac{\nu _{1,\mathrm{rec}}^{3}}{[\exp (h\nu
_{1,\mathrm{rec}}/KT_{\mathrm{rec}})-1]}\simeq \nonumber
\\ \frac{\nu _{1,\mathrm{rec}}^{3}}{[h\nu _{1,\mathrm{rec}}/KT_{\mathrm{rec}}]}= (KT_{\mathrm{rec}}/h)\nu _{1,\mathrm{rec}}^{2}. \label{24}
\end{eqnarray}

At present, expecting the standard cosmological behaviour of cosmological
photon cooling, one as well would assume to have:

\begin{eqnarray}
\varepsilon (\nu _{1,0})\sim \frac{\nu _{1,0}^{3}}{[\exp (h\nu
_{1,0}/KT_{0})-1]}\simeq \frac{\nu _{1,0}^{3}}{[h\nu _{1,0}/KT_{0}]} = \nonumber
\\ (KT_{0}/h)\nu _{1,0}^{2}=(KT_{0}/h)\nu _{1,\mathrm{rec}}^{2}(\frac{c^{2}R_{\mathrm{rec}}}{8\pi GM})^{2}. \label{25}
\end{eqnarray}

From the context derived before this then allows for an identity of the
results to require the following equivalence:

\begin{equation}
(KT_{\mathrm{rec}}/h)\nu _{1,\mathrm{rec}}^{2}=(KT_{0}/h)\nu _{1,\mathrm{rec}}^{2}(\frac{c^{2}R_{\mathrm{rec}}}{8\pi GM})^{2}, \label{26}
\end{equation}

or:

\begin{equation}
T_{\mathrm{rec}}=T_{0}(\frac{c^{2}R_{\mathrm{rec}}}{8\pi GM})^{2}. \label{27}
\end{equation}

Inserting the expected value from standard cosmology:

\begin{equation}
T_{\mathrm{rec}}/T_{0}\simeq 10^{3}=(\frac{c^{2}R_{\mathrm{rec}}}{8\pi GM})^{2} \label{28}
\end{equation}

one then is lead to the following replace value:

\begin{equation}
\frac{R_{\mathrm{rec}}}{M}=\sqrt{10^{3}}\frac{8\pi G}{c^{2}}=95\frac{H_{0}^{2}}{c^{2}\rho _{c}}, \label{29}
\end{equation}

or the result:

\begin{equation}
\frac{1}{800}=\frac{GM}{c^{2}R_{\mathrm{rec}}}. \label{30}
\end{equation}

The findings in section 2 are the basis for the considerations presented in the next section.

\section{Energy density of photons after the
recombination era}
According to the standard model of cosmology the spectral
character of LTE- photons created during the phase of recombination just before the last
photon scattering, say at $z_{\mathrm{r}}\approx 1100$ under conventional CMB\
assumptions, can be described by a Planck distribution with a temperature $%
T_{\mathrm{r}}\approx 3000$\,K. Due to the expansion of the universe the wavelengths of
the free CMB photons are usually assumed to increase and only
because of that the spectrum stays Planckian, since then the Planck
temperature of the CMB\ photons decreases according to $T(z)=(1+z)\cdot T_{0}
$ (see e.g. Fahr \& Zoennchen 2009; Fahr \& Sokaliwska 2015). This
cooling is expected to explain the present-day CMB temperature $%
T(z=0)=T_{0}\approx 2.7$\,K. For a Planck spectrum the energy density $%
\epsilon _{\gamma }$ of the CMB photons and the associated photon number
density $n_{\gamma }$ are given by the well known equations (see Sciama 1971; Peebles 1993):

\begin{equation}
\epsilon_\gamma = \frac{4\sigma_{\mathrm{SB}}}{c}T^4 \approx 0,00472\,\mathrm{eV} \frac{T^4}{\mathrm{cm^3}}, \label{31}
\end{equation}

with $\sigma_{\mathrm{SB}}$ denoting the Stefan-Boltzmann constant
and $T$ given in $K$, and

\begin{equation}
n_{\gamma }=\int_{0}^{\infty }\frac{\varepsilon _{\gamma }(\lambda )\lambda
d\lambda }{hc}\approx 20,28\frac{T^{3}}{\mathrm{cm^{3}}}, \label{32}
\end{equation}

with $h$ the Planck constant and again $T$ given in K. In
this cosmological "standard view" we can then for a present-day CMB\
temperature of $T_{0}=2,7$\,K expect an energy density $\epsilon _{\gamma,0}$
of roughly 

\begin{equation}
\epsilon_{\gamma,0} \approx 0,00472\,\mathrm{eV} \frac{T_{0}^{4}}{\mathrm{cm^3}} \approx 0.26\,\mathrm{eV/cm^{3}} \label{33}
\end{equation}

and a photon number density $n_{\gamma ,0}$

\begin{equation}
n_{\gamma,0 } \approx 20,28\frac{T_{0}^{3}}{\mathrm{cm^{3}}} \approx 400/\mathrm{cm^{3}}. \label{34}
\end{equation}

We now want to apply our new idea of a constant wavelength of
each CMB photon at world times after the recombination era. Since the energy of single CMB\ photons does not change under
the above assumption of a constant wavelength,  the number density of the CMB photons changes as
function of the inverse volume increase due to the expansion, i.e the ratio $%
(S_{r}/S_{0})^{3}$ with $S_{r}=S(t_{r})$ being the scale parameter at worldtime $t_{r}$ at the end of the recombination era and $%
S_{0}$ its reference value at the worldtime $t_{0}$, i.e. today. Thus, the
present photon number density would then again be as in Eq. (\ref{34}):

\begin{equation}
n_{\gamma ,0}^{\lambda =const.}=20,28\frac{T_{\mathrm{r}}^{3}}{\mathrm{cm^{3}}}(\frac{S_{\mathrm{r}}}{S_{0}})^{3}\approx 400/\mathrm{cm^{3}}. \label{35}
\end{equation}

Here $T_{r}$ is the temperature at the end of the
recombination era ($T_{\mathrm{r}}=(1+z_{\mathrm{r}})T_{0}$). The present photon number
density would be in fact the same, since it is now the increase of the scale parameter by a
factor $S_{0}/S_{\mathrm{r}} = (1+z_{\mathrm{r}})\approx 1100$ that reduces the number density to the same value
that further above we had derived for the case of a conventionally
decreasing CMB radiation temperature.

In contrast to that, under the new auspices of a constant wavelength the present-day cosmological
photon energy density would, however, not be identical with the value given
in Eq. (\ref{33}), but for energy-conserving CMB photon propagation now would be higher
by a factor $(1+z_{\mathrm{r}})$ because the photons have not lost energy during the
expansion. Their energy density thus is the one valid at the recombination
era, reduced by the reciprocal of the volume factor. With the assumption of
a constant wavelength the present photon energy density at $S=S_{0}$ can
therefore be written as:

\begin{equation}
\epsilon_{\gamma,0}^{\lambda = const.} = \frac{4\sigma_{\mathrm{SB}}}{c}T_{\mathrm{r}}^4 (\frac{S_{\mathrm{r}}}{S_{0}})^3. \label{36}
\end{equation}

Instead of Eq. (\ref{33}) we then finally obtain: 

\begin{eqnarray}
\epsilon_{\gamma,0}^{\lambda = const.} \approx (1+z_{\mathrm{r}}) 0,26\,\mathrm{eV / cm^3} \approx \nonumber \\ 
286\,\mathrm{eV / cm^3} \approx 286\,\mathrm{MeV / m^3}. \label{37}
\end{eqnarray}

At this point of the paper we like to strictly point out and strongly emphasize the following 2 points:

\begin{enumerate}
	\item The above considerations, namely the scale behavior $\propto S^{-3}$ and the related
		results in the Eqs. (\ref{35}) and (\ref{36}), are valid only for freely propagating CMB photons. 
		Then - instead of $S^{-4}$ - the $S^{-3}$-scaling of the photon energy density must be used for the free CMB photons to calculate the cosmic 
		dynamics in the frame of the Friedmann equations (FLRW cosmology).
	\item As soon as these CMB photons interact, e.g. with an observer, the detected wavelength 
	will reflect the cosmological wavelength shift and show the well-known $S^{-4}$-scaling of the photon energy density due to the reasons explained in Section 2. Thus, an 
	observer at the present time will always see a Planckian CMB spectrum with a temperature of $\approx$ 2.7\,K!
\end{enumerate}

Now it is extremely interesting to recognize that in
comparison to the energy density given in Eq. (\ref{37}) the energy density of visible, baryonic matter in the
universe is about $\rho _\mathrm{{0,B}}\approx 0,3\,m_{\mathrm{proton}}c^{2}/\mathrm{m^{3}}\approx
282\,\mathrm{MeV/m^{3}}$. Interestingly enough, both energy density values are nearly
identical. In other words: the energy density of free CMB photons shows the same drop-off with scale as 
the energy density of the baryonic matter even in the present universe. One
hence can conclude that on the basis of assumptions made in this paper one
would come to the conclusion, that the energy density of free CMB photons at present
is not negligible compared to the matter density but comparable in numbers,
and this is valid not only for the cosmic epoch now, but should stay valid
for the cosmological evolution into the past.

\section{Discussion and Conclusion}
We have based our investigations on the idea of a constant
wavelength of freely propagating photons, i.e. photons that, while
propagating through the universe, do not interact neither with each other
nor with material particles like protons or electrons. The concept of
zero-lapse of proper time for photons in both, STR and GTR, must lead in
 consequence to unaffected physical characteristics of photons propagating
along light geodetics. This means that the wavelength does not continously 
change during the free flight but remains constant. During this phase of free propagation the 
photon energy density simply scales according to $S^{-3}$ (i.e. a pure volume effect) which is then the relevant dependence to be
taken serious for the input into the energy-momentum source tensor $T_\mathrm{{ik}}$ entering the
calculation of the cosmic scale dynamics with
Robertson-Walker-Friedmann-Lema\^{i}tre equations. With other words: the energy 
density of freely propagating cosmic photons (no matter if CMB or galactic photons) is 
changing like the photon number density of these photons according to $S^{-3}$, 
since the energy of each individual photon does not change with the expansion. What happens is,
that the clock of the observer is cosmologically detuned with respect to that of the emitter, i.e. 
photons communicating between emitter and observer are 
differently qualified at their origin and at their absorber (observer).
As soon as an observer interacts with the photons, i.e. analyzes the wavelength or
measures the frequency of them, they appear, as if their energy density had
followed the conventionally believed $S^{-4}$ law, because these
interactions occur at another worldtime compared to that of their origin.
In consequence one must realize that the observed photon energy density 
($\propto S^{-4}$) differs from the energy density of free photons the latter of which, however, is relevant
for the cosmic dynamics ($\propto S^{-3}$).

\end{document}